\documentclass[12pt,preprint]{aastex}

\slugcomment{Article for PASP}

\lefthead{Carl H. Gibson}
\righthead{Primordial Turbulence}

\begin{document}

\title{ Hydro-gravitational fragmentation, diffusion and
condensation of the primordial plasma, dark-matter and gas}

\author{Carl H. Gibson\altaffilmark{1}}
\affil{Departments of Mechanical and Aerospace Engineering and 
Scripps Institution of Oceanography, University of California,
     San Diego, CA 92093-0411}

\email{cgibson@ucsd.edu}

\altaffiltext{1}{Center for Astrophysics and Space Sciences,
UCSD}

\begin{abstract}  

The first structures were proto-voids formed in the primordial
plasma. Viscous and weak turbulence forces balanced
gravitational forces  when the scale of causal connection
$L_H \equiv ct \approx L_{SV} \equiv (\gamma \nu / \rho
G)^{1/2}
\approx L_{ST}
\equiv {\varepsilon}^{1/2}/(\rho G)^{3/4}$  at time
$t
\approx $ 30,000 years ($10^{12}$ s), with
$c$ the speed of light,
$\gamma \approx 1/t$ the rate of strain, and
$\nu$ the kinematic viscosity, where $L_{SV}$ and $L_{ST}$  are
viscous and turbulent Schwarz scales of hydro-gravitational
theory \citep{gib96}.  The photon viscosity
$\nu
\approx 4
\times 10^{26}$ $\rm m^2$ $\rm s^{-1}$  allows only weak
turbulence from the Reynolds number
$Re_H \equiv c^2 t/\nu
\approx 200,
$ with fragmentation at $\rho {L_{SV}^3}
\approx  10^{16} M_{\sun}$ to give proto-supercluster voids,
buoyancy forces, fossil vorticity turbulence, and strong sonic
damping.   The expanding, cooling plasma continued
fragmentation to  proto-galaxy-mass
$\approx  10^{12} M_{\sun}$, with
$\rho
\approx 10^{-17}$ kg $\rm m^{-3}$ and
$\gamma \approx 10^{-12}$ $\rm s^{-1}$ preserved as fossils of
the weak turbulence and first structure. Turbulence
fossilization by self-gravitational buoyancy explains the 
$\delta T/T \approx 10^{-5}$ cosmic microwave background
temperature fluctuations, not sonic oscillations in
cold-dark-matter fragments.  After plasma to gas transition at
$t
\approx $ 300,000 years ($10^{13}$ s), gas fragmentation
occurred within proto-galaxies at $L_J \approx 10^4 L_{SV}$ and
$L_{SV}
\approx L_{ST}$ scales to form
proto-globular-star-cluster (PGCs) clouds of $10^{12}$
 small-planetary-mass primordial-fog-particles
(PFPs). Dark PGC clumps of frozen PFPs persist as
inner-galaxy-halo dark matter, supporting Schild's 1996
quasar-microlensing interpretation. Non-baryonic dark matter,
with
$D \gg 10^{28}$
$\rm m^2$ $\rm s^{-1}$, diffused into the plasma 
proto-cluster-voids and later  fragmented as
outer-galaxy-halos at diffusive Schwarz scales
$L_{SD} \equiv (D^2 /\rho G)^{1/4}$,  indicating $m \approx
10^{-35}$ kg  weakly-collisional
fluid particles.  Observations \citep{gs03} support the
theory. 
\end{abstract}


\section{Introduction}  We consider the hydrodynamic evolution
of the hot big bang expanding universe after mass-energy
equality to determine when gravitational forces were first
able to form structure under the influence of viscous and
turbulent forces.  All flows of plasmas and gases with large 
$Re
\equiv \delta v \times L/\nu \ge Re_{cr}$ are unstable to the
formation of turbulence according to the 1883 Reynolds number
criterion for transition, where $Re_{cr}$ is a finite critical
value of
$Re$ above which laminar flows are impossible, 
$\delta v$ is the velocity difference on scale
$L$ and $\nu$ is the kinematic viscosity of the fluid. From the
first universal similarity hypothesis 
 for turbulence \citep{kol41}, the universal critical Reynolds
number value
$Re_{cr}
\approx 25-100$ applies to the Hubble flow as it does for all
others.  Gluon-neutrino-photon viscosity values $\nu \approx
c^2 t$ before mass-energy equality at 25,000 years give
subcritical
$Re
\approx 1$. Cosmic microwave background anisotropy extended
Self Similarity (ESS) coefficients  \citep{bs02} closely matching
those for high Reynolds number turbulence are attributed to
fossils of pre-inflationary turbulence  \citep{gib01} at
temperatures
$10^{28} \-- 10^{32}$ K too hot for large lepton viscosities
to exist that might otherwise prevent big bang turbulence.  
Predictions of spectral forms and other first and second order
turbulent flow parameters from Kolmogorovian universal
similarity theories for turbulence and turbulent mixing have
been widely validated in numerous atmospheric, oceanic and
laboratory flows and numerous fluids
\cite{gib91}.  No experimental counterexamples exist, either
for the Reynolds number turbulence transition criterion or
Kolmogorovian universal similarity at low order.  Linear
stability theories suggesting the possibility of steady
inviscid flows
\cite{ray80} have been recognized as unreliable for real
fluids since the
\cite{pra21} discovery of viscous instabilities and because
much larger values of
$Re_{cr}$ are predicted by such theories than observed in
laboratory experiments and numerical simulations
\cite{whi91}.  

Prior to the 1993 discovery that the anisotropies
$\delta T/T$ of the cosmic microwave background temperature are
very small ($\approx 10^{-5}$) it was consistently assumed by
all authors that
$Re$ values of the expanding universe would be supercritical
($\gg 100$), so that both the plasma and the subsequent gas
would be strongly turbulent with primordial turbulence the
crucial factor in all subsequent gravitational structure
formation. Density fluctuations produced and mixed by the
turbulence would trigger gravitational collapse to form
structures such as stars and galaxies at mass scales
determined by the primordial  turbulence.   From energy
arguments vol Weizsacker 1951 showed  the Jeans 1902 linear
acoustical criterion for gravitational instability in ideal
fluids fails in strongly turbulent flows. He proposed that
Kolmogorov's incompressible turbulence expression
$\delta v \sim L^{1/3}$ for velocity differences
$\delta v$ between points separated by distances
$L$ should  be used to compute the turbulent kinetic energy of
a possibly unstable gas or plasma cloud, asserting that the
turbulent kinetic energy of the cloud should be less than the
gravitational potential energy as the criterion for
gravitational instability in such clouds. Chandrasekhar 1951
also rejected the Jeans 1902 criterion for the gravitational
instability of strongly turbulent flows but overlooked
Kolmogorov's theory in any form and simply added a turbulence
pressure
$p_T
\sim
\rho(\delta v)^2 $ to the fluid pressure $p$ in the expression
for Jeans's length scale
$L_J$
\begin{equation}  L_J \equiv V_S/(\rho G)^{1/2} 
\approx (p/\rho^2 G)^{1/2} ,
\label{eq1}
\end{equation}
 where $\rho$ is the density, $G$ is Newton's gravitational
constant and
$V_S$ is the sound speed, to give a Chandrasekhar turbulent Jeans scale
$L_{JCT}
\equiv [(p + p_T)/\rho^2 G]^{1/2}$.  Star formation rates in
the cold molecular clouds of the Galaxy disk are about 50
times less than expected from Eq. \ref{eq1}, presumably
because 
$L_{JCT} \ge L \ge L_J$, where $L$ is the size of the cloud,
Scheffler and Elsasser 1988, p438.  Doppler broadened
molecular absorption lines give strong evidence of
Kolmogorovian turbulence in such clouds
\cite{fal90}.  A dissipation rate
$\varepsilon \approx 10^{-6}$ $\rm m^{2} \, s^{-3}$ is
estimated from the third order velocity structure function
measured in the Ursa Major cirrus cloud \cite{miv99}, giving
$L_{ST} \approx 8 \times 10^{18}$ m ($\rho \approx 10^{-19}$
$\rm kg \, m^{-3}$) much larger than the cloud size so that
star formation is prevented (see Table 1).

The hydro-gravitational theory (HGT) of gravitational
structure formation 
\cite{gib96} abandons the Jeans 1902 theory in its entirety; not only for 
strongly turbulent flows but for flows that are weakly
turbulent or nonturbulent. $L_J$ in Eq.
\ref{eq1} should not be interpreted as either the minimum scale or
maximum scale of gravitational instability as proposed by Jeans 1902. 
Such misinterpretations have resulted in the dark matter paradox.
For a self gravitating ideal gas of nearly uniform density
$\rho$ and temperature $T$,
$L_J$ represents the maximum scale of acoustical pressure and
temperature equilibration
$L_{IC} \equiv (RT/\rho G)^{1/2}$, where
$R$ is the gas constant and $p$ is the
pressure (Gibson and Schild 1999ab).  Such a field of nearly
uniform plasma with known properties formed after the big bang and
turned to gas at 300,000 years.  From HGT, non-acoustic density
perturbations in the primordial plasma and gas are absolutely unstable to
structure formation, and viscous or weakly turbulent fluid forces at
$L_{ST}
\approx L_{SV}
\approx L_K
\equiv (\nu^3/\varepsilon)^{1/4} \approx L_H \equiv ct$, or diffusion at
$L_{SD}$ determine the smallest scales of gravitational instability, not
$L_J$, where $L_K$ is the Kolmogorov scale and $L_H$ is the Hubble scale
of causal connection. In the hot
primordial plasma
$L_J \ge L_H$, so by the Jeans 1902 criterion no structure
could form. Cold-dark-matter (CDM) non-baryonic
fluid was invented with small
$L_J$ values to permit  gravitational structure formation
consistent with observations
\cite{pad93}.  However, the necessarily strong diffusivity $D_{CDM}
\gg c^2 t$ of the weakly collisional non-baryonic dark matter
in the plasma epoch prevents its condensation and rules out
CDM models
\cite{gib00} because $(L_{SD})_{CDM} \gg L_H$ in the plasma epoch.  

To correct the Chandrasekhar 1951 expression, the turbulent
pressure 
$\sim
\rho(\delta v)^2
$ should be substituted rather than added to
$p$ in Eq.
\ref{eq1} and the complete Kolmogorov 1941 expression
$\delta v \approx (\varepsilon L)^{1/3}$ should be substituted
for
$\delta v$.  Solving for the critical length scale at which
inertial forces match gravitational forces gives
\begin{equation} L_{ST} \equiv \varepsilon ^{1/2}/(\rho
G)^{3/4},
\label{eq2}
\end{equation} where $L_{ST}$ is defined as the turbulent
Schwarz scale \cite{gib96} and $\varepsilon$ is the viscous
dissipation rate of the turbulence.  

If the turbulence of the primordial plasma flow is weak, as
indicated by the small CMB fluctuations, then viscous forces
$F_V
\approx \rho \nu \gamma L^2$ determine the smallest scale of
gravitational instability, balancing gravitational forces $F_G
\approx
\rho ^2 G L^4$ at the viscous Schwarz scale
$L_{SV}$, where
\begin{equation} L_{SV} \equiv (\nu \gamma /
\rho G)^{1/2},
\label{eq3}
\end{equation}
$\nu$ is the kinematic viscosity of the fluid,
$\gamma$ is the rate of strain, and $\rho$ is the density.  
The turbulent Schwarz scale of Eq. \ref{eq2} is closely
related to the Ozmidov length scale $L_R
\equiv  (\varepsilon /N^3)^{1/2}$ of stably stratified
turbulent flows, where the stratification frequency $N$  has
a physical significance similar to the inverse free fall time
$(\rho G)^{1/2}$ and
$L_R$ is derived by  matching turbulence forces with buoyancy
forces to find the critical length scale.  The viscous Schwarz
scale of Eq. \ref{eq3} near $Re_{cr}$ is analogous to the
buoyancy-inertial-viscous scale
$L_{BIV} \equiv (\nu/N)^{1/2}$ that arises in fossil turbulence
theory
\cite{gib99a}.  Turbulence is strongly inhibited and rapidly
fossilized by buoyancy forces in the ocean and atmosphere at
$L_R$ scales, and astrophysical turbulence is strongly
inhibited and fossilized at
$L_{ST}$ scales in self gravitating fluids.  Because kinetic
and gravitational forces of a flat universe are closely
matched at the horizon scale $L_H$, it follows that whatever
turbulence levels existed at the time of first structure
formation (when, for the first time,
$L_{SV} \approx L_{ST}
\le L_H$) would be rapidly damped by buoyancy forces and the
horizon length, density, mass, and the hydrodynamic
parameter ($\varepsilon$ or $\gamma$) preserved by
hydrodynamic fossils.

Silk and Ames 1972 suggest that the large size of $L_J
\gg L_H \equiv ct$ in the plasma epoch with sound speed
$V_S
\approx c/3^{1/2}$ prevents gravitational condensation of
plasma by the Jeans 1902 criterion.  By their galaxy formation theory,
strong turbulence produced density fluctuations that served as
nuclei for galaxy formation at the time of photon decoupling
when the sound speed $V_S$ dramatically decreased by a factor
of
$3\times10^4$.  Other studies claiming that strong primordial
turbulence should set the scale of galaxies include Gamov 1952,
Ozernoi and Chernin 1968, Ozernoi and Chernin 1969, Oort 1970,
and Ozernoi and Chebyshev 1971.    

All such strong turbulence theories of structure formation were
rendered moot by the 1993 measurements of very small
temperature fluctuations
$\delta T/T
\approx 10^{-5} \approx \delta v / v \approx
\delta \rho / \rho \approx \delta p / p
\approx \delta a/ a$ in the cosmic microwave radiation (CMB)
data from the 1989 COsmic Background Explorer (COBE)
satellite,   rather than values of
$\delta v / v \approx 10^{-1} \-- 10^{-2}$ that would result
from fully developed turbulence, where $a$ is the cosmic scale
factor and $\delta (T,v,\rho,p,a)$ represent fluctuation
magnitudes. A subcritical horizon scale Reynolds number
$Re_H
\equiv c^2 t/\nu \le 10$ at the time
$10^{13}$ s of plasma-gas transition requires an enormous
kinematic viscosity
$\nu
\ge  10^{29}$ $\rm m^2$ $\rm s^{-1}$ to be subcritical, much
larger than $\nu \approx 10^{25}$ m$^2$ $\rm s^{-1}$ estimated for the
primordial plasma then \citep{gib00}. For a terrestrial comparison, the
kinematic viscosity of the Earth's upper mantle is $\nu
\approx 10^{21}$ $\rm m^2$ $\rm s^{-1}$ from glacial rebound
rates (Professor Robert Parker of SIO, personal
communication).  Implicitly it has been assumed in the
astrophysics literature after these COBE observations that the
Hubble flow of the expanding universe must somehow be
intrinsically stable to turbulence formation, independent of
Reynolds number.    Textbooks on structure formation in the
universe such as Padmanabhan 1993 make no mention of viscosity,
diffusivity, turbulence, or Reynolds number in their
discussions of the process.  No reference in the literature
has been found that attempts to justify this implicit (and unwarrented)
assumption.  An example of strong turbulence generated by the Hubble flow
is shown in Figure 1.  Powerful Hubble flow drag forces separate
protosuperclusters, protoclusters, and protogalaxies as they form by
gravitational fragmentation in the primordial plasma and early gas epochs
according to HGT.  Hubble flow galaxy Reynolds numbers of order $10^{12}$
shown in Figure 1 have decreased to values $\approx 10^{4}$
or less at present.

The assumption made by CDM
hierarchical clustering cosmology models  (CDMHCCs) that the
Hubble flow is stable to the formation of turbulence is inconsistent with
the universal similarity theory of turbulence, which is the basis of HGT. 
Strong turbulence in the plasma epoch is ruled out from HGT by the small
values
$\delta T/T \approx 10^{-5}$ of the CMB observations, so buoyancy forces
resulting from gravitational structure formation must have dominated the
damping of turbulence because viscous forces are
inadequate and no other fluid forces exist.  Turbulent transition cannot
fail by lack of triggering perturbations since 
$\delta T/T$ fluctuations are observed at scales $L > ct$ in
the CMB that can nucleate growth of  vorticity and structure
once they enter the horizon.  Neither can it be argued that a
lack of time prevents self gravitational or fluid mechanical
nonlinearity.  Once
$L_{K}
\le L_H$ at turbulence transition the eddy overturn time is
$t$. The decreasing viscous stresses in the baryonic component
permit fragmentation of supercluster to galaxy masses in the
plasma epoch so that the hierarchical clustering of
subgalactic scale CDM halos to form these structures in the gas
epoch is unnecessary, even if such small CDM halos were
physically possible (they are not).  Because the nonbaryonic
dark matter is necessarily strongly diffusive, such small CDM
halos are excluded by HGT 
\citep{gib00}.  Observations of galaxy-QSO correlations and
discordant cluster red shifts rule out CDMHCCs 
\citep{gs03}.  CDMHCCs are also excluded by observed density
distributions near galaxy cluster cores that fail to
match universal forms computed by numerical simulation
\citep{sa02}.

In the following
$\S$\ref{sec1} we consider whether an inviscid expanding
universe is stable or unstable to the formation of turbulence.
If it is unstable according to the conventional Reynolds
number criterion, what constraints on viscosities and
structure formation in the plasma epoch can be inferred from
observed CMB anisotropies?  We then examine the hydrodynamic
parameters and structures to be expected from the
\citep{gib96} nonlinear gravitational structure formation
theory during the plasma epoch, in
$\S$\ref{sec3}, and in the early gas epoch, in
$\S$\ref{sec4}.  Conclusions are summarized in
$\S$\ref{sec5}.

\section{The absolute instability of inviscid flows}
\label{sec1}

The instability of expanding flows is discussed in
$\S 23$ of Landau and Lifshitz 1959.  The equations of
momentum conservation in a fluid may be written
\begin{equation}
\partial \vec v / \partial t = -\nabla B +
\vec v
\times
\vec \omega + \nu \nabla ^2 \vec v  +
\vec F_M + ...
\label{eq4}
\end{equation}  where $B \equiv p/\rho + v^2 /2 + \phi$ is the
Bernoulli group of mechanical energy terms,
$\vec
\omega
\equiv \nabla
\times \vec v $ is the vorticity, $\vec v
\times
\vec \omega$ is the inertial vortex force that causes
turbulence,
$\nu \nabla ^2 \vec v$ is the viscous force that damps it out,
$\vec F_G = - \nabla \phi$ is the gravitational force and has
been absorbed in
$B$,
$\phi$ is the gravitational potential energy per unit mass in
the expression
$\nabla ^2 \phi = 4 \pi \rho G$, $G$ is Newton's constant,
$\vec F_M$ is the magnetic force, and other forces have been
neglected.  Eq.
\ref{eq4} applies in a gas or plasma when a sufficient number
of particles are assembled, so that the particle separation
$L_P$ and the collision distance $L_C$ are much smaller than
the size
$L$ of the assemblage or the scale of causal connection $L_H
\equiv ct$, where $c$ is the speed of light and $t$ is the age
of the universe.  Turbulence develops whenever the
inertial-vortex force of the flow is larger than the other
terms; that is, if the Reynolds number 
$Re \equiv (\vec v\times \vec \omega) / ( \nu
\nabla ^2 \vec v) $, Froude number $Fr
\equiv (\vec v\times \vec \omega) / \vec F_G$, and all other
such dimensionless groups exceed critical values.   

In  Landau and Lifshitz 1959 $\S 23$ an exact solution of  Eq.
\ref{eq4} for an incompressible viscous fluid attributed to G.
Hamel 1916 (usually termed the Jeffrey-Hamel flow, White 1991)
gives multiple maxima and minima for expanding flows between
plates. This solution is used to illustrate the relative
instability of expanding flows compared to converging flows.  
The Jeffrey-Hamel converging flow solution approaches the
solution for converging ideal fluid flow and thus might appear
to be stable to the formation of turbulence at high Reynolds
number because the turbulent intensity
$\delta v /v$ decreases along a streamline as
$v$ increases.  Converging sections are used in wind and water
tunnels before test sections to decrease the turbulent
intensity, but the turbulent viscous dissipation rate
$\varepsilon$ and turbulent velocities actually increase in
such flows
\citep{bat53}.  Can steady inviscid flows of any kind be
stable?

What about the stability of the expanding universe which is not
incompressible but is a uniform expansion with rate-of-strain
$\gamma
\approx 1/t$, where $\gamma$ is generally termed the ``Hubble
constant'' and the expansion is termed the ``Hubble flow''? 
Instead of decreasing along a streamline with
$1/x$ as for the incompressible diverging Jeffrey-Hamel flow,
the speed
$ v \approx \gamma x$ increases with distance
$x$.  Does this mean the expanding Hubble flow is stable,
similar to the converging Jeffrey-Hamel flow where the speed
also increases with distance?  Does this mean that the small
CMB temperature anisotropies simply reflect the fundamental
stability of a Hubble flow, and does not imply that large
viscous or buoyancy forces must have been present in the
plasma epoch?  Are self gravitating fluids  fundamentally
different from stratified natural fluids in that the first
turbulence of the Hubble flow is caused by gravitational
forces rather than inhibited by them as in stratified flows?  

According to the further analysis and discussion in Landau and
Lifshitz 1959, in
$\S 27$ titled ``The onset of turbulence'', steady inviscid
flows are absolutely unstable.  Thus, all flows should develop
turbulence at high enough Reynolds numbers, including the
diverging Hubble flow of the expanding universe. In their
derivation, maximum amplitudes of Fourier modes
$|A|_{max}  \sim (Re - Re_{crit})^{1/2}$ are expressed as
functions of their departures from critical Reynolds numbers
$Re_{crit}$ and it is shown that the individual modes grow to
finite values with increasing Reynolds number, but with an ever
increasing number of modes as $Re \rightarrow
\infty$.  Landau-Lifshitz admit that prediction of the mode
amplitudes is mathematically difficult and that such stability
analysis has had limited success in predicting the transition
to turbulence except to confirm the 1883 Reynolds criterion,
for which there is no experimental counterexample.  As we have
seen, in apparent counterexamples such as the Jeffrey-Hamel
converging incompressible flow the increasing velocity along
streamlines masks the developing turbulence, but does not
prevent it.

The absolute instability of steady inviscid flows can be
understood from the first two terms of Eq. \ref{eq4}, shown in
Eq.
\ref{eq5}.  Such a flow must be irrotational to remain steady
with
$\partial
\vec v /
\partial t = 0$ and $B$ constant.  Otherwise the vorticity
$\vec\omega$ would produce inertial vortex forces $\vec v
\times \vec \omega $ that would spread the rotational region
indefinitely to larger and smaller scales by undamped turbulent
diffusion.  If a variation in speed occurs along one of the
steamlines, then accelerations
\begin{equation}
\partial \vec v / \partial t = -\nabla B 
\label{eq5}
\end{equation} develop that amplify any perturbations in $v$
with increasing time. Increasing
$v$ requires increases in both $B$ and its gradient, and
decreasing
$v$ decreases both
$B$ and its gradient.  From Eq.
\ref{eq5}, positive speed perturbations increase $-\nabla B$
and cause speed increases, and negative speed perturbations
cause decreases in
$-\nabla B$ and cause speed decreases.  Vorticity $\vec \omega
> 0$ develops and forms turbulence, which will grow in size and
kinetic energy.  This positive feedback is independent of the
continuity equation or the equation of state for the fluid. 
Finite length scale perturbations of any of the hydrophysical
parameters ($v,p,\rho$) in a steady, inviscid, irrotational
flow will cause local perturbations in the vorticity on the
same finite scale, with resulting formation and growth of
turbulent inertial vortex forces
$\vec v \times \vec \omega$ and thus turbulence at larger and
smaller scales, drawing energy from the assumed variations of
$v$ along streamlines.  Even the extreme case of steady flow is
unstable to a vorticity perturbation, since the rotational
region of the
$\vec\omega$ perturbation without viscous damping will spread
its vorticity and kinetic energy, and thus turbulence, to
indefinitely larger volumes by turbulent diffusion.

We conclude that steady inviscid flows are absolutely unstable,
confirming the 1959 Landau-Lifshitz result and the conventional
Reynolds criterion for turbulence formation. Viscosity is not
necessary to the formation of turbulence, only its evolution. 
From the vorticity conservation equation following a fluid
particle in a fluid with variable density
\begin{equation} D \vec \omega / Dt = \partial \vec \omega
/\partial t + (\vec v
\cdot
\nabla )  \vec \omega  = \vec \omega \cdot
\vec{\vec e} + (\nabla \rho
\times \nabla p) / \rho ^2 + \nu \nabla ^2 \vec
\omega
\end{equation} we see variations in the density of the fluid
can produce vorticity if pressure and density gradients are not
aligned, at rate
$(\nabla \rho
\times \nabla p) / \rho ^2
$, leading to unconstrained inertial vortex forces
$\vec v \times \vec
\omega$ and thus turbulence.  Vorticity is produced by vortex
stretching at a rate $\vec \omega \cdot
\vec{\vec e}$, where $\vec{\vec e}$ is the rate of strain
tensor. Turbulence is defined as an eddylike state of fluid
motion where the inertial vortex forces of the eddies are
larger than any other forces that tend to damp the eddies out
\citep{gib99a}.  Turbulence always starts at the smallest
possible scale permitted by viscous forces, and cascades to
larger scales by a process of eddy pairing and entrainment by
the turbulence of irrotational fluid
\citep{gib91}.  Fourier modal analysis fails to properly
describe the formation of turbulence, gravitational structure
formation, or small scale turbulent mixing at small Prandtl
numbers.  These failures result from sacrificing realistic
physical models for mathematical convenience by considering
the linear behavior of sine waves rather than the nonlinear
behavior of finite-scale local perturbations
\citep{gib96}.  

What about cosmic drag?  It is sometimes argued that
turbulence is prevented by the expansion of the universe
because momentum decreases as
$V(t)=V_0 / a(t)$ from general relativity, where $V_0$ is an
initial velocity perturbation and $a(t)$ is the cosmic scale
factor which monotonically increases with time $t$ as the
universe expands.  Although the momentum and velocity of a
perturbation may decrease, the proper length scale of the
perturbation $L(t) = L_0 a(t)$ will increase, so that $a(t)$ in
the Reynolds number $Re(t) \equiv VL/\nu$ will cancel.  To
first order, the Reynolds number after inflation and before
mass-energy equivalence is
$Re \approx 1$ because
$V \approx c$, $L \approx ct$, and $\nu \approx c^2 t$. 
Before inflation, much larger Reynolds numbers were possible
\citep{gib00}.

It is not true that simply because the initial perturbation of
a nonlinear process is small that the process can be accurately
described by linear theories.  In particular, just because
remnant density perturbations
$\delta
\rho / \rho \approx 10^{-5}$ from big bang quantum
gravitational chaos are small does not mean that their
evolution can be accurately described by linear methods once
they reenter the horizon.  Decreasing the size of
$\delta
\rho / \rho$ by a factor of $10^{-5}$ increases the
gravitational condensation time by less than a factor of two. 
Cold dark matter theories that suggest an acoustic peak in the
CMB temperature spectrum are therefore questionable.  The
gravitational response to density perturbations is always
nonlinear and requires nonlinear fluid mechanics for its
description independent of the size of the perturbation or the
predictions of the linear, acoustic theory of Jeans 1902.

What about energy conservation?  Won't pressure support or
thermal support prevent gravitational condensation at scales
smaller than
$L_J$?  Won't continued gravitational collapse require a loss
of thermal energy to prevent pressure stabilization, and won't
this require a spontaneous and highly efficient flow of heat
from a cold object into a hot environment?   These
misconceptions are all part of Jeans's 1902 legacy.  Consider
a volume of initially stagnant, constant density gas, smaller
than the horizon, with mass perturbation
$M'$ suddenly placed near its center. This system is absolutely
unstable to gravitational condensation or void formation,
depending on whether $M'$ is positive or negative. 
Gravitational acceleration starts immediately with radial
velocity $v_r \approx -tGM'(t)/r^2$, and
mass flux
$ 4
\pi r^2  \rho v_r{(t)} \approx  4
\pi \rho G M'(t) t = dM'(t)/dt$ independent of radius.  Thus $M'(t) =
M'(0) exp (2 \pi \rho G t^2) =  M'(0) exp [2 \pi
(t/{\tau_G})^2]$.    The density,
temperature, and dynamical pressure
$p/\rho + v^2 /2$ remain constant during the gravitational free fall
process except in the small space-time region of the nonacoustic density
nucleus at $r \ll L_J$ and
$t \approx \tau_G \equiv (\rho G)^{-1/2}$ (Gibson 1999a, Gibson $\&$
Schild 1999a).  Everything happens at once when $t \rightarrow  \tau_G$. 
Since it takes $ t \approx \tau_G$ for information to propagate a
distance $L_J$, no pressure support mechanism is possible to prevent the
self gravitational collapse or void formation at nonacoustic density
perturbations.

In any real fluid, the Hubble flow is unstable at all scales
where the Reynolds number exceeds a universal value
$Re_{crit}
\approx 100$.  Thus, $Re \approx \delta v
\times x / \nu \approx \gamma x^2 / \nu
\approx 100$ at a critical length scale
$x_{crit}
\approx 10 (\nu / \gamma)^{1/2}$.  The viscous dissipation
rate $
\varepsilon \approx
\nu \gamma ^2$, so
\begin{equation}  x_{crit} \approx  10 ({\nu} ^{3/2}
/\varepsilon^{1/2})^{1/2}  \approx  10 L_K,
\label{eq6}
\end{equation} where 
\begin{equation} L_K \equiv (\nu^{3}/\varepsilon)^{1/4} =
(\nu/\gamma)^{1/2}
\label{eq7}
\end{equation} is the 1941 Kolmogorov length scale.  Turbulence
always begins at scales of
$\approx 10 L_K$ and is inhibited at smaller scales by viscous
forces.  These small eddies pair, pairs of eddies pair with
other eddy pairs, and so forth.  Irrotational (and therefore
nonturbulent) fluid is entrained into the interstices of the
turbulent domain as ideal flows, is made turbulent at
Kolmogorov scales by viscous forces, and supplies the kinetic
energy of the turbulence.  We now use these results to examine
the formation of turbulence, and its inhibition, during the
plasma epoch before
$10^{13}$ s (300,000 years).

\section{The plasma epoch}
\label{sec3}

What about the formation of turbulence in the plasma epoch? 
Since the Hubble flow is unstable to the formation of
turbulence, either viscous forces or buoyancy forces, or both,
must have been present to prevent strong turbulence.  When did
the first turbulence form?  What was the viscosity of the
plasma required to prevent turbulence?

From COBE to WMAP, numerous experiments have been undertaken to
resolve the small scale fluctuations of the CMB. Super-horizon
contributions to the
$\delta T$ variance are approximately constant with a
Sachs-Wolfe plateau of about
$2 \times 10^{-5}$ K for angular separations
$\theta$ greater than about 1-2 degrees corresponding to the 
horizon scale
$L_H
\approx 3 \times 10^{21}$ m existing at this plasma-gas
transition time $10^{13}$ s
\citep{lin99}.  From measurements at smaller sub-horizon
scales a sonic, or doppler, peak of about 
$8
\times 10^{-5}$ K at $\theta \approx 0.5$ degrees and
smaller-amplitude, smaller-scale,  harmonics are attributed to
undamped sound waves in the plasma sloshing in CDM clump
potential wells in  the  gravitational potential.  

This sonic peak explanation of the CMB is questionable for at
least three reasons:  1. the postulated CDM fluid with
$L_{SD} > L_H$ (see Eq. \ref{eq11} below) is too diffusive to
condense;  2.  no sound source of any kind exists, and
certainly not the non-turbulent super-powerful sound source
that would be required to match the observations; 
3.  even if a super-powerful source of sound could be
identified, the sound would be rapidly damped by viscous
forces because the sonic attenuation coefficient
$\alpha  \approx
\nu/V_S \lambda^2$ is $\gg \lambda^{-1}$ since
$\nu
\approx V_S L_C$ and $L_C \gg \lambda$ for all the relevant
sonic wavelengths
$\lambda$.  

Reason 3. is why sonic fluctuations of temperature in the
relatively noisy atmosphere of the earth rarely exceed the 1
db reference level
$\delta T/T
\approx 10^{-10}$, Pierce and Berthelot 1990, and why whales
near Japan can be heard from California but eagles cannot. Time
$t_{FS}
\approx 10^{12}$ s ($30,000$ years) is indicated as
the time of first structure formation since this is the time
when the increasing horizon mass
$\rho (ct)^{3}$ just matches the observed mass of superclusters
$\approx 10^{46}$ kg
\citep{gib97b}.  This supercluster mass is
$10^{-6}$ times the present horizon mass
$(ct)^3 \rho_{crit} = 10^{52}$ kg since the observed supervoid
size is $10^{-2} \times L_H$.   The observed globular star
cluster density 
$\approx 10^{-17}$ kg $\rm m^{-3}$ just matches the baryonic
density existing at $t \approx 10^{12}$ s, also indicating
$t_{FS}
\approx 10^{12}$ s as the time when the plasma first began
fragmentation.  Voids formed at that time should expand for a
brief period as rarefaction waves with velocities limited by
the sound speed
$V_S = c/3^{1/2}$, giving a structural rather than sonic
peak in the range $0.6 >
\theta_{SP} > 0.1$ degrees, as observed, with a monotonic
decrease of the
$\delta T$ power spectrum reflecting fragmentation to galactic
scales, possibly with acoustical harmonics from the
rarefaction waves. Further fragmentation at smaller and
smaller scales limits the amplitude of
$\delta \rho /
\rho$ to small values as $M_{SV}$ decreases toward proto-galaxy
masses in the cooling, expanding plasma.  

To prevent turbulence at the horizon scale at
decoupling requires a viscosity
$\nu_{crit} \approx c^2 t/100 \approx 10^{28}$
$\rm m^2$ s$^{-1}$, which is too large for the baryonic
component by any known mechanism.  Setting
$x_{crit} = L_H = 10 L_K$ in Eqs. \ref{eq6} and
\ref{eq7} with $\gamma = 1/t$ gives a value of
$\nu = (ct/100)^2 \gamma = 9 \times 10^{26}$
$\rm m^2$
$\rm s^{-1}$ for our estimated $t_{FS}
\approx 10^{12}$ s.  This large value of
$\nu$ is only slightly larger than that required to prevent
turbulence at the time of first structure.  Once
gravitational structure formation begins, buoyancy forces
will inhibit  turbulence.

Densities were larger at this earlier time ($30,000$ yr) so
mean free paths for collisions $L_C \approx (\sigma n)^{-1}$
were shorter, where $\sigma$ is the collision cross section and
$n$ is the particle density. The physical mechanism of viscous
stress in the plasma epoch is photon collisions with the free
electrons of the plasma (Silk \& Ames 1972, Thomas 1930).  The
electrons then drag along the protons and alpha particles of
the primordial plasma to maintain electrical neutrality.  The
kinematic viscosity is then
\begin{equation}
\nu \approx L_C \times v = c/\sigma_T n_e
\label{eq9}
\end{equation} where $\sigma_T = 6.65 \times 10^{-29}$ $\rm
m^2$ is the Thomson cross section for scattering and
$n_e$ is the number density of the free electrons.
Substituting $n_e
\approx 10^{10}$ $\rm m^{-3}$ for the electron number density
at
$t = 10^{12}$ s \citep{win72} gives $\nu \approx 4 \times
10^{26}$ $\rm m^2$
$\rm s^{-1}$, which is close to our estimated minimum
$\nu$ value required to inhibit turbulence.  The collision
distance
$L_C \approx 1.5 \times 10^{18}$ m is less than the horizon
scale
$L_H = 3 \times 10^{20}$ m, so the assumption of collisional
fluid dynamics in Eq. \ref{eq9} is justified.  The viscous
dissipation rate $\varepsilon
\approx
\nu \gamma ^2 \approx 4 \times 10^{-2}$ $\rm m^2$ $\rm s^{-3}$
gives a Kolmogorov scale
$L_K \approx 2 \times 10^{20}$ m from Eq.
\ref{eq7}.  Since $10 L_K \ge L_H$, the Hubble flow of plasma
should be viscous and laminar or weakly turbulent.

The baryonic density at $t \approx 10^{12}$ s was $\rho \approx
2 \times 10^{-17}$ kg $\rm m^{-3}$
\citep{win72}.  The strain rate at turbulence fossilization was
$10^{-12}$
$\rm s^{-1}$.  Thus, from Eq.
\ref{eq3}
\begin{equation} L_{SV} \approx (10^{-12} \times 4 \times
10^{26} / 2 \times 10^{-17} \times 6.672 \times
10^{-11})^{1/2} \approx 5 \times 10^{20} \, \rm m,
\label{eq8}
\end{equation}  approximately matching the horizon scale
$L_H = 3 \times 10^{20}$ m.  The horizon scale baryonic mass
$M_H
\equiv L_H^3
\times
\rho = (ct)^3 \rho = 5 \times 10^{44}$ kg is close to the
baryonic mass of superclusters
$(M_{SC}\approx 10^{46}$ kg includes the non-baryonic
component), so these are suggested as the first structures of
the universe, formed by fragmentation when the viscous Schwarz
scale first matches the horizon scale, Gibson 1996, 1997ab,
1999ab. 

Proto-superclusters formed by fragmentation rather than
condensation because void formation is augmented by the
expansion of the universe but condensation is inhibited.  Thus
proto-supercluster-voids expand in the plasma epoch while the
proto-superclusters between these voids also grow, but more
slowly, by internal fragmentation, preserving the density of
the fragments.  Further fragmentation at
$L_{SV}$ scales down to protogalaxy masses with little change
in the baryonic density due to fossil density turbulence
formation is proposed  by HGT
\citep{gib96}. Turbulence formation is
inhibited at every stage of the plasma epoch by a combination
of viscous and buoyancy forces, and there is no energy source
for sound other than the gravitational void formation.
Temperature fluctuations observed in the CMB are proposed as
fossils of big bang turbulence and
fossils of the first structure formation, Gibson 2000. 

In contrast, Silk 1989 Fig. 10.1 traces the evolution of an
adiabatic galaxy mass pressure fluctuation as it drops below
the Jeans mass at a redshift of $z
\approx 10^8$ and oscillates as an undamped sound wave in the
necessarily inviscid plasma epoch with
$10^4$ density contrast until decoupling at $z
\approx 10^3$.  It seems unlikely that any such loud sounds
($\ge 100$ db) could start at that time, less than a week
after the big bang.  If somehow they were started they
would be rapidly damped, within another week, by the large
photon viscosity, not to mention damping by the expansion of
the universe (cosmic drag).  Sonic pressure fluctuations
$p
\approx p_o exp[-\alpha x]$, where $p_o$ is the initial
pressure, $x$ is the direction of propagation.  The sonic
attenuation coefficient $\alpha \approx
\nu
\omega^2/V_S^3 = \nu /V_S \lambda ^2$, where the frequency
$\omega = V_S/\lambda$, $\lambda$ is the wavelength, and $V_S$
is the sound speed
\citep{pie90}.  Thus,
$p/p_o
\approx exp [-(\nu/V_S \lambda ^2) x ]
\ll 1$ for distance $x \ge \lambda$ if
$\nu \ge V_S \lambda$, and this will be true since $\nu$
increases with time as the universe density decreases and
$\lambda \le ct$ is limited in size by the time $t$ when the
sound wave was created. 

What about the non-baryonic dark matter (NB) required to make
up the critical density of a flat universe?  Its cross section
$\sigma_C$ for collisions with ordinary matter must be very
small or it would have been detected based on the expression
$\sigma_{C} = m_p(GM/r)^{1/2}/\rho D_{NB}$, where
$m_p$ is the particle mass and $D_{NB}$ is the diffusivity
inferred from outer-halo dimensions $r$ of galaxies or clusters
of mass
$M$ \citep{gib00}.  Thus such material must have large mean
free paths for collisions and large diffusivities
$D_{NB}
\equiv L_{NB}
\times v_{NB}$ compared to $D_{B}$ for baryonic matter since
$L_{NB} \gg L_B$ and
$v_{NB}
\approx v_B$.  From measurements of the mass profile of Abell
1689 by Tyson \& Fischer 1995, Gibson 1999b estimates the
non-baryonic dark matter of the dense galaxy cluster is
$D_{NB} \approx 10^{28}$
$\rm m^2$
$\rm s^{-1}$ by setting the radius of curvature of the profile
to
$L_{SD}$. Since neutrinos are now known to have mass, an
obvious non-baryonic candidate is neutrinos, which have
densities comparable to the density of photons and very small
cross sections for collisions since they interact with
baryonic matter mostly through the weak force.  Large numbers
of neutrinos were formed in nucleosynthesis and their unknown
number of flavors and abilities to convert between flavors
leaves their total mass a mystery.  Assuming a neutrino
collision cross section of
$\sigma_n \approx 10^{-40}$ $\rm m^2$ and number density $n_n
\approx 10^{20}$
$\rm m^{-3}$ gives a mean free path $L_n$ of
$10^{20}$ m, so collisional dynamics apply.  Cross sections for
light ($\approx 10^{-35}$ kg) particles like neutrinos give
such small
$\sigma$ values, but $10^{-25}$ kg particles like neutralinos
give
$\sigma \approx 10^{-22}$
$\rm m^2$, much larger than $\approx 10^{-46}$
$\rm m^2$ theoretical values or the $\le 10^{-42}$
$\rm m^2$  values excluded by experiments
\citep {gib00}.

In the case of strongly diffusive matter in weakly turbulent
flows, gravitational condensation is limited by a match between
the diffusion velocity of an isodensity surface
$V_D \approx D/L$ and the gravitational free fall velocity
$V_G \approx L/\tau_G$, giving the diffusive Schwarz scale
\begin{equation} L_{SD} \equiv (D^2/\rho G)^{1/4}
\label{eq11}
\end{equation} where the diffusivity $D_n \approx L_n \times c
\approx 3 \times 10^{28}$ $\rm m^2$ $\rm s^{-1}$.  This gives
$L_{SD} \approx 10^{21}$ m during the plasma epoch, much larger
than any of the structures formed and larger than the horizon
for part of the epoch.  Any such nonbaryonic material would
diffuse away from the protogalaxies and proto-superclusters as
they fragment, to fill the voids between.  Non-baryonic
materials  fragment as the last stage of gravitational
structure formation to form protosuperhalos when the baryonic
protosuperclusters separate by scales larger than
$L_{SD}$.  This is contrary to cold dark matter models that
require CDM condensation as the first rather than last stage of
structure formation, producing, rather than being produced by,
the baryonic structure. 

The necessary condition for the diffusive Schwarz scale 
$L_{SD}$ of  Eq.
\ref{eq11} to determine the minimum scale of gravitational
condensation is
\begin{equation} D \ge \nu \gamma \tau_G
\label{eq12}
\end{equation} for viscous flows.  Since
$\gamma \tau_G \ge 1$ and $D \approx
\nu$ for baryonic matter, the scale
$L_{SD}$ only applies to nonbaryonic matter.  Substituting
$D_n \approx 3 \times 10^{28}$ $\rm m^2$ $\rm s^{-1}$ and $\rho 
\approx 10^{-23}$ kg $\rm m^{-3}$ for the density of
a galaxy cluster gives $L_{SD} \approx 3 \times 10^{22}$ m
(Mpc) as the scale for gravitational fragmentation of the
non-baryonic dark-matter halo of a small
galaxy cluster with total mass $ \approx 4
\times 10^{44}$ kg.

Thus a proper description of structure formation in the
primordial self-gravitational fluids of the early universe
requires more than the linearized Euler equation with gravity
and the density equation without diffusion or gravity, as
assumed by
\citep{jns02}.  All the forces in the momentum Eq.
\ref{eq4} are needed except ($\vec F_M + ...
$).  The appropriate non-acoustic density conservation
equation near density maxima and minima is 
\begin{equation}
\partial \rho / \partial t + \vec v \cdot
\nabla \rho = D_{eff}\nabla ^2 \rho , 
\label{eq13}
\end{equation} where $ D_{eff} \equiv D - L^2 / \tau_G$ and
$D$ is the molecular diffusivity of the density; that is, on
scales
$L \le L_J$ so that the pressure adjusts rapidly, and on
scales $L
\ge {L_{SX}}_{max}$ (the maximum Schwarz scale) where gravity
dominates fluid forces and molecular diffusion
\citep{gib99b}.  The density
$\rho$ depends on temperature and species concentration
variations and their diffusivities, and not simply the
pressure as assumed by Jeans.  The problem is similar to the
turbulent mixing problem
\citep{gib68} except for the remarkable fact that for clouds
of fluid with sizes $L_J \ge L
\ge {L_{SX}}_{max}$, gravitational diffusivity takes over and
the effective diffusivity
$D_{eff}$ becomes negative.  Thus, rather than reaching a local
equilibrium between local straining and diffusion at the
Batchelor length scale $L_{B} \equiv (D/\gamma)^{1/2}$ near
density extrema as in turbulent mixing theory with a monotonic
decrease toward ambient values, densities in the
self-gravitational fluids of astrophysics increase to large
values or decrease toward zero at these points due to
gravitational instability
\citep{gs99a}.

\section{The gas epoch}
\label{sec4}

From standard cosmology and the CMB
observations, the initial conditions of the
gas epoch are precisely defined.  Little or
no turbulence was present, as discussed
previously, so the rate of strain of the
fluid was larger than
$\gamma \approx 1/t
\approx 10^{-13}$ $\rm s^{-1}$ existing at
that time and smaller than the fossil
vorticity turbulence value in the structures
$\gamma_{FS} \approx 10^{-12}$
$\rm s^{-1}$. The density of the protogalaxies
cannot have been much different from the
fossilized initial fragmentation density
$\rho_{FS}
\approx 10^{-17}$ kg $\rm m^{-3}$ since there
was insufficient time for collapse.  The
temperature at decoupling was
$T_o \approx 3\,000$ K.  The composition was
75\% H and 25\% He by mass.  Therefore the
kinematic viscosity of the primordial gas was
about $3 \times 10^{12}$ $\rm m^2$ $\rm
s^{-1}$, from
$\mu  \equiv \rho \times \nu$ in standard gas
tables with a weighted average $\mu (T_o)$,
with gas constant $R$ about $3\,612$ $\rm m^2$
$\rm s^{-2}$ $\rm K^{-1}$,
\cite{gib99b}.  Viscous dissipation rates were
only $\varepsilon \approx \nu \gamma ^2
\approx 3 \times 10^{-14}$ $\rm
m^2$ $\rm s^{-3}$ so the Kolmogorov scale
$L_K \approx 5 \times 10^{12}$ m from Eq.
\ref{eq8}  was a factor of $5 \times 10^{8}$
smaller than the horizon scale.

Fragmentation of the neutral gas protogalaxies occurred
simultaneously at both the Jeans scale
$L_J$ of Eq. \ref{eq1} and the viscous Schwarz scale
$L_{SV}$ of Eq. \ref{eq3}, where for the primordial gas
conditions
$L_J \approx 10^{4} L_{SV}
\gg L_{SV}$.  The physical mechanism of this Jeans scale
fragmentation is not the mechanism proposed by Jeans 1902. 
Temperatures in growing voids at scales smaller than
$L_J
\approx (RT/\rho G)^{1/2}$ adjust by particle diffusion to
remain constant at
$T = p/\rho R$ as the gravity driven rarefaction waves of void
formation propagate, where the term ``void'' indicates a
density deficiency rather than $\rho = 0$.  As the density
decreases the pressure decreases.  Particle speeds and
temperatures are constant as long as the particle diffusion
time
$\tau_P \equiv L/(RT)^{1/2}$ is less than the gravitational
free-fall time
$\tau_G$; that is, for scales
$L
\le L_J$.  For scales
$L$ larger than
$L_J$ the diffusion time $\tau_P$ is larger than $\tau_G$,
causing temperatures in these large voids to decrease as the
voids grow because particle diffusion cannot maintain constant
temperature and acoustical equilibrium.  When this happens,
radiation heat transfer from the warmer surroundings increases
the temperature, and thus also the pressure, within the voids,
and the increased pressure  accelerates the void formation,
isolating blobs of gas at some multiple of the Jeans scale to
form PGCs.

Substituting the 
$T$ and 
$\rho$ values of the primordial gas gives  
$L_J
\approx  5 \times 10^{17}$ m, and $M_J
\equiv L_J^3 \rho \approx  10^{35 \-- 36}$ kg.  Substituting
$\rho$,
$\gamma$ and
$\nu$ values in $L_{SV} \equiv (\nu
\gamma / \rho G)^{1/2}$ gives $L_{SV} \approx
 10^{14}$ m and $M_{SV} \equiv L_{SV}^3
\rho \approx 10^{24 \-- 25}$ kg, a factor of $\approx 10^{12}$
smaller than
$M_J$.  The Jeans scale objects are called
``Proto-Globular-Clusters'' (PGCs) and the 
$L_{SV}$ scale objects are called ``Primordial Fog
Particles'' (PFPs).  From the observational evidence it appears
that many if not most PGCs have not yet dispersed and most of
their PFPs have not yet accreted to form stars, so that both
persist as the dominant component of galactic baryonic dark
matter
\citep{gib96}. The calculated masses of PGCs and PFPs depend on
universal proportionality constants of order one that
will emerge from observations.  Observations of globular star
clusters indicate a mass $10^{5 \-- 6} M_{\sun}$
matching our calculated PGC value of $10^{35
\-- 36}$ kg and densities close to the fossilized initial fragmentation density
$\rho_{FS}
\approx 10^{-17}$ kg $\rm m^{-3}$. The calculated PFP
mass
$10^{24 \-- 25}$ kg matches observations of
$\approx 10^{-6} M_{\sun}$ ``rogue planets'' by Schild 1996 as
the dominant component of the lensing galaxy in a lensed
quasar system. Evidence supporting HGT has recently been summarized
\citep{gs03}, and includes the appearance of PFP candidates brought out
of cold storage by evaporation near hot objects such
as white dwarfs in planetary nebula.  Figure 2 shows PFP candidates in
the Helix planetary nebula which is the one closest to earth,
photographed by the Hubble Space telescope.  Thousands of cometary
globules appear with mass values, densities, and separation distances as
predicted by HGT.

\section{Conclusions}
\label{sec5} We conclude that the small amplitude $\delta T/T
\approx 10^{-5}$ of measured temperature fluctuations in the
cosmic background radiation is evidence of strong turbulence
damping by both a photon viscosity $\nu
\approx 4
\times 10^{26}$ $\rm m^2$ $\rm s^{-1}$ and buoyancy forces of
viscous-gravitational structure formation beginning
approximately
$30,000$ years after the Big Bang. 

The hypothesis  is rejected that the small CMB 
fluctuations reflect hydrodynamic stability
of the Hubble flow.   This would  require a critical Hubble
flow Reynolds number of
$Re_{cr_H}
\approx 10^5$, contrary to universal similarity hypotheses of
Kolmogorov 1941 for turbulence and strong experimental
evidence that the universal critical Reynolds number of
transition is
$Re_{cr} \le 25$.  Steady inviscid flows are absolutely
unstable to the formation of turbulence, as shown in
$\S$2 and as derived by Landau \& Lifshitz 1959.  Buoyancy
forces from gravitational structure formation in the plasma
epoch are therefore required to explain the lack of turbulence
at the time of plasma to gas transition.

The hypothesis  is rejected that the CMB
spectral peak with
$L \approx 0.4 L_H$ is a doppler or sound horizon
with
$L
\approx V_S t$ because no source of sound exists to produce the
observed
$\delta T/T
\approx 8
\times 10^{-5}$ peak value, and because strong viscous damping
in the plasma epoch would rapidly flatten any such sonic
peaks.  Persistent sonic oscillations of baryonic matter
sloshing in CDM potential wells as a sound source is rejected
because no CDM potential wells are possible in the plasma
epoch, because viscous damping would occur, and because recent
strong observational evidence excludes CDMHCC scenarios,
Gibson \& Schild 2003.  Instead, the observed spectral peak at
scales
$L \approx ct$ is interpreted as evidence of the first
hydro-gravitational structure formation. Secondary acoustic
peaks observed may reflect rarefaction wave oscillations of
hydro-gravitationally driven proto-supercluster void formation
near sonic velocities.  Hydro-gravitational theory suggests the
first structures to form were proto-supercluster-voids at the
viscous Schwarz scale
$L_{SV}$, when $L_{SV} > L_H$ first matched the increasing
horizon scale
$L_H$. Rapid expansion of the universe during the plasma epoch
prevented gravitational condensation but enhanced void
formation,
$\S$3.    

As shown in $\S$4, fragmentation of the primordial gas 
occurred simultaneously at
$L_J$ and
$L_{SV}$ scales to form proto-globular-clusters (PGCs) and
primordial-fog-particles (PFPs).  Estimated PGC masses match
the observed globular star cluster masses of
$10^6 M_{\sun}$ and estimated PFP masses match the observed
``rogue planet'' dark matter masses of
$10^{-6} M_{\sun}$ in lensed quasars
\citep{sch96}.  Most PGCs and their PFPs are observed to
persist as dark clumps of frozen planetoids, forming the
dominant component of  $\approx 100 kpc = 3 \times
10^{21}$ m galactic dark-matter inner-halos
\citep{gs03}, with non-baryonic dark matter fragmenting to form
outer galactic dark-matter halos at
$L_{SD} \approx Mpc = 3 \times 10^{22}$ m scales.

All evidence suggests the early universe was an extremely
gentle place, with practically no turbulence or sound anywhere
after the big bang and prior to the formation of stars. 
Buoyancy and large photon viscosities damped sound and
turbulence in the plasma epoch.   Gravitational condensation
formed PFPs and prevented turbulence in the early stages of
the gas epoch.  As the universe continued to expand and cool,
some of these small-planetary-mass objects experienced an
accretional cascade to larger mass-scales to form
the first very small stars.  This cascade 
was a gentle process to produce the
remarkably spherical distributions of these long-lived,
tightly-packed stars in globular star clusters and the more
numerous dark or dim PGC-PFP
baryonic-dark-matter structures with the same
$\rho \approx 10^{-17}$ $\rm kg \, m^{-3}$ density  and the PFP
mass $\approx 10^{24}$ kg preserved as fossils of
the weak turbulence and large density at the time of first
structure
\citep{gib00}.

\acknowledgments

The author is grateful for many constructive questions and
comments from Rudolph Schild.  This paper is dedicated to the
memory of George Keith Batchelor, 8 March 1920 $\--$ 30 March
2000.

\begin{figure}
         \epsscale{0.6}
         \plotone{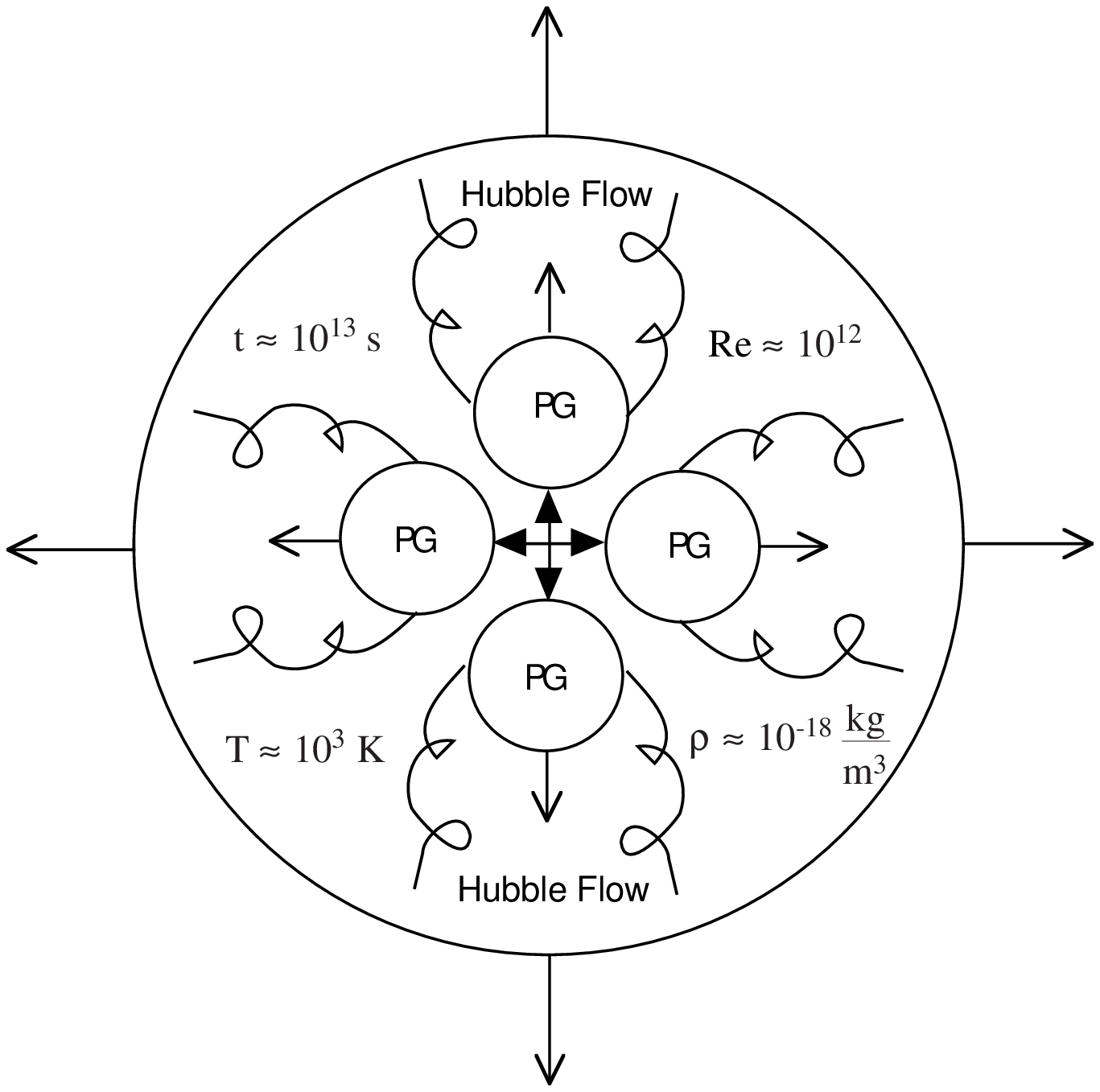}
         \caption{Turbulent Hubble flow wakes of proto-galaxies (labeled
PG) bound by gravity (dark double arrows) in a proto-cluster, soon after
the plasma to gas transition at time $t \approx 10^{13}$ s (300,000 years)
after the big bang.  From HGT \citep{gib96} the PG size is $
\approx 10^{20}$ meters, with primordial H-He mass $ \approx 10^{42}$ kg
($\approx 10^{12} M_{\sun}$) and density $\rho \approx 10^{-18}$ kg $\rm
m^{-3}$.  The Hubble flow velocity $v_H \equiv r \times \gamma_H
\approx 10^{7}$ m s$^{-1}$ at the radius $r$ of the PGs.  Hubble flow
drag forces will separate the protogalaxies, as it separated the
protoclusters and protosuperclusters also formed in the plasma epoch by
gravitational fragmentation.  Nonbaryonic dark matter moves freely
through the galaxies by diffusion and the Hubble flow to fill the voids. 
The horizon scale
$L_H$ where
$v_H = c$ is
$3
\times 10^{21}$ m (100 kpc), 10 times the size of the outer sphere shown.}
         \end{figure}

\begin{figure}
         \epsscale{0.7}
         \plotone{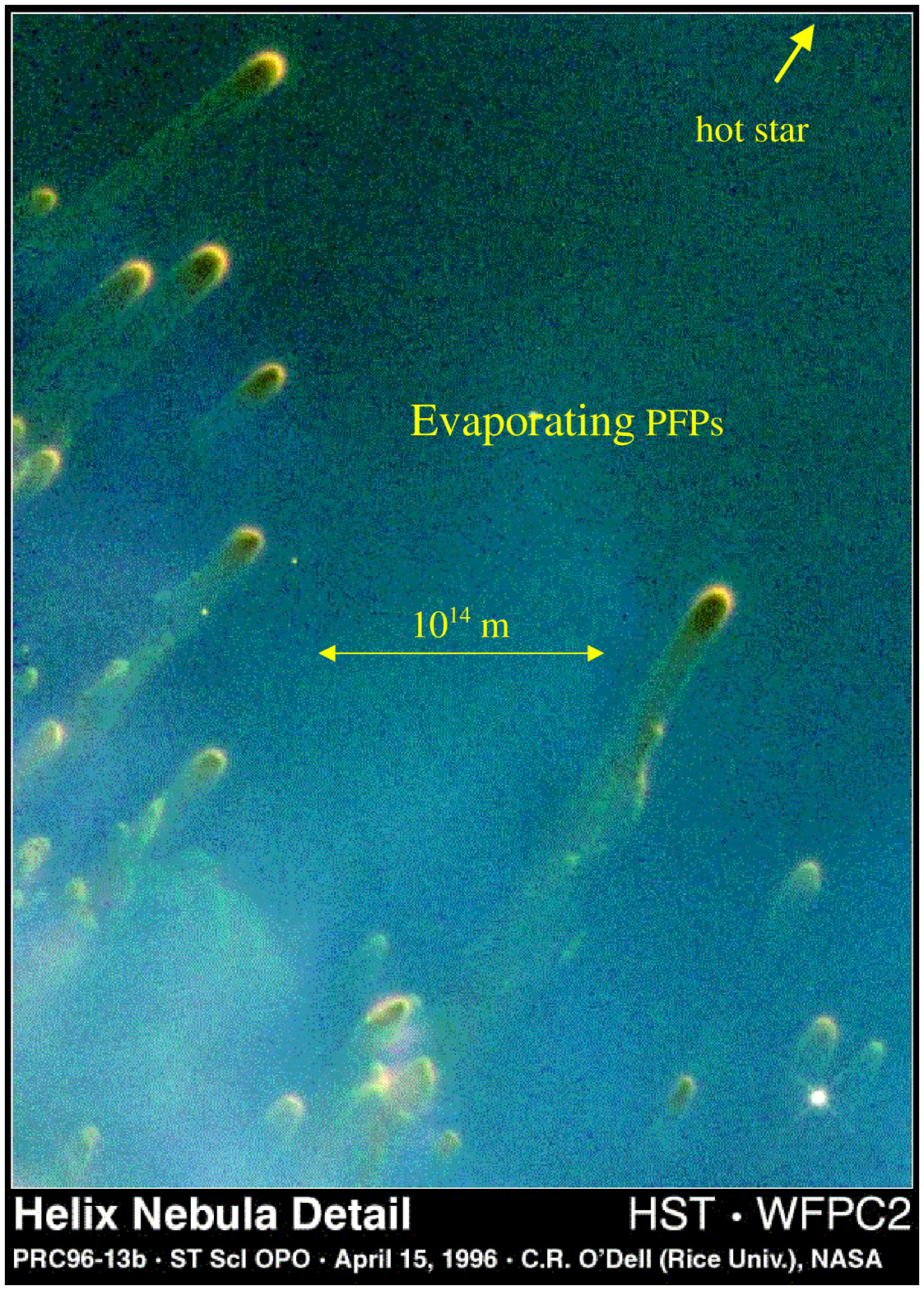}
         \caption{PFP-like objects \citep{gs03} observed by HST in the
Helix Planetary Nebula,
$4.5
\times 10^{18}$ m from Earth
\citep{ode96}.  Evaporation of the  frozen H-He objects produces $10^{25}$ kg
photo-ionized cocoons, with H-wakes pointing away from the hot ($\approx 
50,000 K$) White Dwarf.  The indicated density of the PN halo is $\rho_{Halo}
\approx M_{PFP} L_{Sep.}^{-3} \approx 10^{-17}$ kg m$^{-3}$, where the PFP
separation distance $L_{Sep.} \approx 10^{14}$ m.  This matches
the baryonic density at the time of first structure $10^{12}$ s (30,000
years) as a fossil of this time of first structure formation as predicted
by HGT.}
         \end{figure}

\begin{deluxetable}{lrrrrcrrrrr}
\tablewidth{0pt}
\tablecaption{Length scales of self-gravitational structure formation}
\tablehead{
\colhead{Length scale name}& \colhead{Symbol}           &
\colhead{Definition$^a$}      &
\colhead{Physical significance$^b$}           }
\startdata Jeans Acoustic & $L_J$ &$V_S /[\rho G]^{1/2}$& ideal gas pressure
equilibration

\\ Chandrasekhar Turbulent & $L_{JCT}$ & $[(p + p_T)/\rho^2
G]^{1/2}$ & turbulence balances gravitation

\\ Schwarz Diffusive & $L_{SD}$&$[D^2 /\rho G]^{1/4}$& $V_D$ balances $V_{G}$
\\  Schwarz Viscous & $L_{SV}$&$[\gamma \nu /\rho G]^{1/2}$& viscous force
balances gravitational force
  \\ Schwarz Turbulent & $L_{ST}$&$\varepsilon ^{1/2}/ [\rho G]^{3/4}$&
turbulence balances gravitation
\\

Kolmogorov Viscous & $L_{K}$&$ [\nu ^3/ \varepsilon]^{1/4}$& turbulence
force  balances viscous force
\\

Batchelor Diffusive & $L_{B}$&$ [D / \gamma]^{1/2}$&
diffusion balances strain rate
\\

Collision & $L_{C}$&$ m \sigma ^{-1} \rho ^{-1}$& distance between particle
collisions
\\

Horizon, Hubble & $L_{H}$&$ ct$& maximum scale of causal connection
\\


\enddata
\tablenotetext{a}{$V_S$ is sound speed, $\rho$ is density, $G$ is Newton's
constant, $p$ is pressure, $p_T \equiv \rho {(\delta v)}^2$, $v$ is
velocity,
$D$ is the diffusivity,
$V_D
\equiv D/L$ is the diffusive velocity at scale $L$, $V_G \equiv L[\rho
G]^{1/2}$ is the gravitational velocity,
$\gamma$ is the strain rate,
$\nu$ is the kinematic viscosity,
$\varepsilon$ is the viscous dissipation rate, $m$ is the particle mass,
$\sigma$ is the collision cross section, light speed $c$, age of universe
$t$.}

\tablenotetext{b}{Magnetic and other forces (besides viscous and turbulence)
are negligible for the epoch of primordial self-gravitational structure
formation considered here
\citep{gib96}.}


\end{deluxetable}

\end{document}